\documentclass[prb,
twocolumn,
superscriptaddress,showpacs,amsmath,amssymb]{revtex4}

\begin{document}

\title{Analog of gravitational anomaly in topological chiral superconductors}

\author{G.E.~Volovik}
\affiliation{Low Temperature Laboratory, Aalto University,  P.O. Box 15100, FI-00076 Aalto, Finland}
\affiliation{Landau Institute for Theoretical Physics, acad. Semyonov av., 1a, 142432,
Chernogolovka, Russia}

\date{\today}

\begin{abstract}
 It is known that the contribution of torsion to the equation for the chiral Weyl fermions can be equivalently considered
in terms of the axial $U(1)$ gauge field. In this scenario the gravitational field transforms to the $U(1)$ gauge field. Here we show that in chiral superconductors the opposite scenario takes place:  the electromagnetic $U(1)$ field serves as the spin connection for the Bogoliubov fermionic quasiparticles. As a result the electromagnetic field gives rise to the gravitational anomaly, which contains the extra factor $1/3$ in the corresponding Adler-Bell-Jackiw equation as compared with the conventional chiral anomaly.
We also consider the gravitational anomaly produced in neutral Weyl superfluids by the analog of the gravitational instanton, the process  of creation and annihilation of the 3D topological objects -- hopfions. The gravitational instanton leads to creation of the chiral charge.
\end{abstract}

\maketitle

\section{Introduction}

Topological materials with Weyl fermions provide the possibility to study quantum anomalies, such as the Adler-Bell-Jackiw (ABJ) chiral anomaly.\cite{Adler1969,BellJackiw1969,Adler2005}  The condensed matter analog of the ABJ  anomaly has been experimentally probed in the Weyl superfluid -- the chiral A-phase of liquid $^3$He.\cite{Bevan1997} In this electrically neutral matter, the chiral anomaly is caused by the effective electromagnetic field produced by deformations, which lead to the the space-time dependence of the superfluid order parameter. Recent discussion of anomalies in Weyl materials can be found in Ref.\cite{Burkov2021}.  Reviews on chiral anomaly effects in the heavy-ion collisions see in Refs.\cite{Kharzeev2021,Mitkin2021}.

The  gravitational anomalies have been also discussed in Weyl materials. 
These anomalies are produced by the effective gravitational fields acting on Weyl fermions. These fields are represented by tetrads, spin connection  and torsion fields.  In particular, the analog of the Nieh-Yan anomaly in terms of torsion\cite{NiehYan1982a,NiehYan1982b,Nieh2007,Obukhov1997} has been developed.\cite{NissinenVolovik2019,NissinenVolovik2020,Nissinen2020,Laurila2020,Stone2020,Stone2020b,Ojanen2020}

 Here we consider the chiral superconductors and show that the $U(1)$ electromagnetic field plays the role of spin connection in the effective tetrad gravity. As a result, these superconductors experience  the analog of gravitational anomaly. As distinct from the conventional chiral anomaly, the corresponding Adler-Bell-Jackiw equation contains  the extra factor $1/3$.
  
\section{From gauge field to spin connection}

The Hamiltonian for Bogoliubov quasiparticles in the $p$-wave homogeneous superfluid $^3$He-A
in a simplest model, which neglects Fermi-liquid corrections, has the following form\cite{VollhardtBook} 
\begin{equation}
H({\bf p}) = \left(\begin{array}{cc}\frac{p^2}{2m} - \mu_0  &{\bf e}_1\cdot {\bf p} + i{\bf e}_2\cdot {\bf p} \\
{\bf e}_1\cdot {\bf p} - i{\bf e}_2\cdot {\bf p}&-\frac{p^2}{2m} + \mu_0
  \end{array}\right)  \,.
\label{1}
\end{equation}
Here ${\bf e}_1$ and ${\bf e}_2$ are the real vectors satisfying conditions ${\bf e}_1\cdot{\bf e}_2=0$, ${\bf e}_1^2= {\bf e}_2^2$.
For simplicity we do not consider the spin structure, i.e. our quasiparticles are considered as spinless. The Hamiltonian (\ref{1}) describes the homogeneous system, and $\mu_0= p_0^2/2m$ is the equilibrium chemical potential. 

This Hamiltonian contains two Weyl points -- topologically stable point nodes in the energy spectrum, which represent monopoles in the Berry phase in momentum space\cite{Volovik1987} (classification in terms of two different topological invariants in the interacting systems can be found in Ref.\cite{Zubkov2021}).
The nodes are at ${\bf p}_\pm=\pm p_0 \hat{\bf l}$, where  the unit vector $\hat{\bf l}\parallel {\bf e}_1\times {\bf e}_2$. The space and time dependence of the Weyl points in the inhomogeneous superfluid produces the effective electromagnetic field  ${\bf A}^{\rm eff}({\bf r},t)=p_0\hat{\bf l}({\bf r},t)$, which acts on the Weyl fermions (the pseudo electormagnetic field in terminology of Ref.\cite{Grushin2020}):
\begin{equation}
H({\bf p}) = e^i_a \tau^a\left(p_i - A^{\rm eff}_i\right) \,.
\label{Weyl}
\end{equation}
Here $\tau^a$ are Pauli matrices in the particle-hole space; and $e^i_a=({\bf e}_1,{\bf e}_2, {\bf e}_3=v_F \hat{\bf l})$ are the space components of the gravitational tetrads, where $v_F=p_0/m$ is the Fermi velocity in the normal Fermi liquid.

 The motion of the Bogoliubov quasiparticles in this effective field gives rise to the analog of the chiral anomaly -- the anomalous nonconservation of the chiral current
 \cite{Adler1969,BellJackiw1969,Adler2005}.  For $^3$He-A the ABJ equation has the following  form:
  \begin{eqnarray}
 \partial_\mu J^\mu_5= \frac{1}{32\pi^2} e^{\mu\nu\rho\sigma}F^{\rm eff}_{\mu\nu}F^{\rm eff}_{\rho\sigma}\,.
 \label{ABJ}
\end{eqnarray}
The prefactor in the ABJ Eq.(\ref{ABJ}) has been confirmed in the experiments on vortex dynamics \cite{Bevan1997}, where the space and time dependence of  the effective gauge field ${\bf A}^{\rm eff}({\bf r},t)=p_0\hat{\bf l}({\bf r},t)$ has been created by moving continuous vortices -- skyrmions in the $\hat{\bf l}$ field.

Let us now consider the electrically charged $p$-wave superfluid, i.e. the superconductor with the same order parameter, and consider the role of the real electromagnetic field in the chiral anomaly. Here we are interested in the influence of the electromagnetic field only, and neglect the effective electromagnetic field, $F^{\rm eff}_{\mu\nu}=0$, i.e. the order parameter is assumed to be fixed.

The inverse Green's function of Bogoliubov quasiparticles in the external  electromagnetic field is:
\begin{equation}
G^{-1}=  -i\partial_t +\tau_3 qA_0({\bf r},t) + H({\bf p} + \tau_3q{\bf A}({\bf r},t)) \,,
\label{GreensFunction}
\end{equation}
where $A_\mu$ is the vector potential of electromagnetic field. This field acts in the opposite ways on particle and hole and hole components of Bogoliubov spinor, that is why the vector potential is accompanied by the matrix $\tau_3$. Here $q=-1$ is electric charge.

Since $\tau_3 =   \frac{1}{2i} (\tau_1 \tau_2 - \tau_2 \tau_1)$, the Green's function can be rewritten in terms of the "covariant derivative": 
\begin{equation}
 D_i = \partial_i + \frac{1}{8} C^{ab}_i (\tau_a \tau_b -  \tau_b \tau_a) ~~,~~ D_t = \partial_t + \frac{1}{8} C^{ab}_0 (\tau_a \tau_b -  \tau_b \tau_a) \,,
 \label{CovariantDerivative}
\end{equation}
where $C^{ab}_\alpha$ are the elements of spin connection, see e.g. Ref. \cite{Laurila2020}.
Let us show that in our case the nonzero components of spin connection are expressed in terms of  the vector potential of electromagnetic field:
\begin{equation}
  C^{12}_i =-C^{21}_i =2A_i({\bf r},t) \,\,,\,\, C^{12}_0 =-C^{21}_0 =2A_0({\bf r},t)\,.
 \label{SpinConnection}
\end{equation}

Expanding the Green's function (\ref{GreensFunction}) near the Weyl node at $p_0\hat{\bf l}$, one obtains
\begin{eqnarray} 
G^{-1} =   -i\partial_t +\tau_3\left(A_0 + \frac{{\bf A}^2}{2m}\right)
\nonumber
\\
+ \tau_3  {\bf e}_3\cdot ({\bf p}-{\bf A}\tau_3-p_0\hat{\bf l})  
\nonumber
\\
+ \frac{1}{2} \left(\tau_1 {\bf e}_1\cdot ({\bf p}-{\bf A}\tau_3-p_0\hat{\bf l})  + {\bf e}_1\cdot ({\bf p}-{\bf A}\tau_3-p_0\hat{\bf l})   \tau_1 \right)
\nonumber
\\
+ \frac{1}{2} \left(\tau_2 {\bf e}_2\cdot ({\bf p}-{\bf A}\tau_3-p_0\hat{\bf l})  + {\bf e}_2\cdot ({\bf p}-{\bf A}\tau_3-p_0\hat{\bf l})   \tau_2 \right)
\nonumber
\\
+ \frac{1}{m}{\bf A}\cdot ({\bf p}-{\bf A}\tau_3-p_0\hat{\bf l})  
\nonumber
\\
+\frac{({\bf p}-p_0\hat{\bf l})^2}{2m}
\,.
\label{FermionHamiltonianAtNodes}
\end{eqnarray}
This expansion is not gauge invariant, because we omitted the gradients of the order parameter. When they are included the gauge invariance is restored.

The quadratic term $({\bf p}-p_0\hat{\bf l})^2/2m$ can be neglected. Then
the resulting Lagrangian can be expressed in terms of effective tetrads $e^\mu_a$, which in our case are constant fields except  for the vector $e^i_0=A^i/m$. This element may produce only the higher order terms in anomaly equation and thus can be neglected. The relevant nonzero elements of the effective spin connection are  $C_\alpha^{12}=-C_\alpha^{21}=({\bf A},A_0)/2$ in Eq.(\ref{SpinConnection}). 
As we already mentioned, the effective vector potential  ${\bf A}^{\rm eff}=p_0\hat{\bf l}$ is kept constant and thus does not contribute to the anomaly.

\section{Gravitational anomaly from electromagnetic field}

Since we neglect the effective gauge field,  $F^{\rm eff}_{\mu\nu}=0$, the chiral anomaly comes only from the curvature of the effective gravitational field. For a single Weyl node one has the following equation for the gravitational anomaly (see e.g. Ref.\cite{Parrikar2014}):
 \begin{eqnarray}
 \partial_\mu J^\mu_5 = \frac{1}{768\pi^2} e^{\mu\nu\rho\sigma}R^{ab}_{\mu\nu}R^{cd}_{\rho\sigma}\eta_{ad}\eta_{bc}\,,
 \label{GravitationalAnomaly}
\end{eqnarray}
where the curvature is:
\begin{equation}
 R^{ab}_{\mu\nu}=\nabla_\mu C^{ab}_{\nu}- \nabla_\nu C^{ab}_{\mu} +  
(C^{ac}_{\mu} C^{db}_{\nu}-C^{ac}_{\nu} C^{db}_{\mu})\eta_{cd}\,.
 \label{curvature}
\end{equation}
According to Eq.(\ref{SpinConnection}) the nonzero elements of the curvature tensor are
\begin{equation}
R_{12\mu\nu}=- R_{21\mu\nu}= \nabla_\mu C^{12}_\nu  -  \nabla_\nu C^{12}_\mu\,.
\label{CurvatureTerms}
\end{equation}
Since the elements of spin connections are expressed in terms of the vector potential of the electromagnetic field, the components of curvature tensor are expressed in terms of external electric and magnetic fields:
\begin{equation}
R_{12\mu\nu}= 2(\partial_\mu A_\nu -\partial_\nu A_\mu)=2F_{\mu\nu}
\,.
\label{CurvatureFieldGauge}
\end{equation}

Then from the anomaly equation (\ref{GravitationalAnomaly}) one obtains:
\begin{eqnarray}
 \partial_\mu J^\mu_5= \frac{1}{384\pi^2} e^{\mu\nu\rho\sigma}R^{12}_{\mu\nu}R^{21}_{\rho\sigma}
= \frac{1}{96\pi^2} e^{\mu\nu\rho\sigma}F_{\mu\nu}F_{\rho\sigma} =
\\
= \frac{1}{3}\, \frac{1}{32\pi^2} e^{\mu\nu\rho\sigma}F_{\mu\nu}F_{\rho\sigma}\,,
 \label{GravitationalAnomaly2}
\end{eqnarray}
or
\begin{equation}
 \partial_\mu J^\mu_5= \frac{1}{24\pi^2} {\bf E}\cdot{\bf B} = \frac{1}{3} \frac{1}{8\pi^2} {\bf E}\cdot{\bf B}\,.
 \label{GravitationalAnomaly3}
\end{equation}
This means that in the superconducting states with Weyl points, the gravitational anomaly becomes the gauge anomaly, but with  the extra factor $1/3$ when compared with ABJ equation (\ref{ABJ}) coming from the effective gauge field  ${\bf A}^{\rm eff}=p_0\hat{\bf l}$.

This factor $1/3$ may have relation to the consistent anomaly \cite{Grushin2020},
and also to the factor $1/3$  obtained in Ref. \cite{Stone2021} for the electromagnetic response and $\theta$-term in the gapped topological superconductors.\cite{Witten2013,Stone2016,Vayrynen2011}

\section{Gravitational anomaly in neutral chiral superfluid and creation of hopfions}

In the electrically neutral superfluids, such as superfluid $^3$He-A, the vector potential of external field is substituted by superfluid velocity: 
${\bf A}\rightarrow m{\bf v}_s$, and the gravitational anomaly becomes
\begin{equation}
 \partial_\mu J^\mu_5= \frac{m^2}{24\pi^2}\,  \partial_t{\bf v}_s\cdot(\nabla\times {\bf v}_s)\,.
 \label{GravitationalAnomalySuperfluid}
\end{equation}
Such anomaly can be produced by the instanton, which represents the process of creation or annihilation of the 3D skyrmions -- 
hopfions in the vector filed $\hat{\bf l}$.\cite{VolovikMineev1977} Hopfions are described by the $\pi_3(S^2)=Z$  topological charge -- the Hopf invariant.\cite{FaddeevNiemi1997}
Recent papers on hopfions in condensed matter can be found in Ref.\cite{hopfion2020}.

In superfluid $^3$He-A, the density of the hopfion  topological charge is expressed in terms 
of the helicity of the superfluid velocity \cite{VolovikMineev1977}:
\begin{equation}
n^0_H= \frac{m^2}{4\pi^2}   ({\bf v}_s\cdot(\nabla\times{\bf v}_s)) \,\,,\, 
N_H=  \int d^3r \,n^0_H\,.
 \label{HopfionCharge}
\end{equation}
Introducing the current density of the topological charge:
\begin{equation}
{\bf n}_H= \frac{m^2}{4\pi^2}   ({\bf v}_s\times\partial_t{\bf v}_s)) \,,
 \label{HopfionChargeCurrent}
\end{equation}
one obtains the (non)conservation  law for the topological charge:
\begin{equation}
 \partial_\mu n^\mu_H= \frac{m^2}{2\pi^2} (\partial_t{\bf v}_s\cdot(\nabla\times{\bf v}_s))\,.
 \label{Conservation}
\end{equation}
The process of the change of the topological charge represents the $\pi_3$ instanton: 
$\partial_\mu n^\mu_H=\delta(t)\delta({\bf r})$. In high energy physics this is the gravitational instanton,
or the so-called torsion vortex.\cite{HansonRegge,AuriaRegge1982}

Then from the anomaly equation (\ref{GravitationalAnomalySuperfluid}) one obtains the connection between the creation of the chiral charge and the creation of the hopfion:
\begin{equation}
 \partial_\mu J^\mu_5= \frac{1}{6}  \partial_\mu n^\mu_H\,.
 \label{GravitationalAnomalyInstanton}
\end{equation}
Here we added the factor 2 to take into account the contribution of two Weyl point into the gravitational anomaly.
Eq.(\ref{GravitationalAnomalyInstanton}) means that due to the gravitational anomaly, the gravitational instanton process of creation of single hopfion is accompanied by creation of 6 chiral fermions. This is the gravitational analog of the Kuzmin-Rubakov-Shaposhnikov scenario of the anomalous electroweak baryogenesis.\cite{Kuzmin1985}

\section{Higher values of topological invariants}

Let us consider the gravitational anomaly in case of the higher order Weyl points. 	The Hamiltonian in Ref.\cite{KonyshevVolovik1988} for Weyl points with topological charges $N$ and $-N$ in chiral superfluids/superconductors has the following form:
\begin{equation}
H({\bf p}) = \left(\begin{array}{cc}\frac{p^2}{2m} - \mu_0  &({\bf e}_1\cdot {\bf p} + i{\bf e}_2\cdot {\bf p})^N \\
({\bf e}_1\cdot {\bf p} - i{\bf e}_2\cdot {\bf p})^N&-\frac{p^2}{2m} + \mu_0
  \end{array}\right)  \,.
\label{N}
\end{equation}
The corresponding spectrum of Bogoliubov quasiparticles near the Weyl points  
\begin{equation}
E^2(\tilde{\bf p}) =(g^{\perp ik} \tilde p_i \tilde p_k)^N + g^{\parallel ik} \tilde p_i  \tilde p_k \,,
\label{spectrum}
\end{equation}
where $\tilde{\bf p}={\bf p} \mp p_0 \hat{\bf l}$ is the momentu counted from the Weyl points, and the elements of the effective metric tensor are:
\begin{equation}
g^{\perp ik} =e_1^i e_1^k  + e_2^i e_2^k = g^{ik} -  g^{\parallel ik} \,\,, \,\,  g^{\parallel ik} =e_3^i e_3^k \,.
\label{g}
\end{equation}
For corresponding superconductors the Hamiltonian is:
\begin{eqnarray}
H({\bf p}) =\tau^3 {\bf e}_3\cdot(\tilde{\bf p} - \tau^3 {\bf A}) +
\nonumber
\\
+(\tau^1-i\tau^2)\left(({\bf e}_1 + i{\bf e}_2)\cdot (\tilde{\bf p} - \tau^3 {\bf A})\right)^N 
\nonumber  
\\
+(\tau^1+i\tau^2)\left(({\bf e}_1 - i{\bf e}_2)\cdot (\tilde{\bf p} - \tau^3 {\bf A})\right)^N.
\label{NGamma}
\end{eqnarray}

The equation (\ref{NGamma}) represents the analog of the Ho$\check{\rm r}$ava gravity.\cite{Horava2009}
Without the first term this is  the $2+1$ Ho$\check{\rm r}$ava gravity, which takes place in graphene, see Ref. \cite{VolovikZubkov2014}. 
With the first term it becomes the anisotropic extension of  the $3+1$ Ho$\check{\rm r}$ava gravity.
While the relativistic invariance is missing, the tetrad contribution to topology and anomaly is still valid. As before, the gravitational anomaly  gives rise to the $U(1)$ anomaly in the external electromagnetic field, again with the extra factor $1/3$:
 \begin{equation}
 \partial_\mu J^\mu_5
= \frac{N}{96\pi^2} e^{\mu\nu\rho\sigma}F_{\mu\nu}F_{\rho\sigma} \,.
 \label{GravitationalAnomalyN}
\end{equation}

For the electric current one has the same expression
 \begin{equation}
 \partial_\mu J^\mu
= \frac{N}{96\pi^2} e^{\mu\nu\rho\sigma}F_{\mu\nu}F_{\rho\sigma} \,,
 \label{GravitationalAnomalyN}
\end{equation}
but the creation of the electric charge is cancelled by the anomaly at the opposite Weyl point with topological charge $-N$.
The electric charge created at one node is annihilated at the opposite node.

The chiral currents for two nodes are added to give the total chiral current
 \begin{equation}
 \partial_\mu J^\mu_{5{\rm tot}}
= \frac{N}{48\pi^2} e^{\mu\nu\rho\sigma}F_{\mu\nu}F_{\rho\sigma} = \frac{1}{3} \frac{N}{16\pi^2}   e^{\mu\nu\rho\sigma}F_{\mu\nu}F_{\rho\sigma} \,.
 \label{ConsistentAnomaly}
\end{equation}

\section{Conclusion}

We demonstrated that in the chiral superconductors with Weyl fermions  the external electromagnetic field serves as the spin connection for the Bogoliubov fermionic quasiparticles. As a result the electromagnetic field gives rise to the gravitational anomaly, which is described   by the Adler-Bell-Jackiw equation with an extra factor $1/3$ compared with the ABJ equation for the conventional chiral anomaly. 

In neutral chiral superfluids  this gravitational anomaly takes place during creation and annihilation of the 3D topological objects -- hopfions. The process analogous to the gravitational instanton leads to creation of the chiral charge. This is the gravitational analog of the Kuzmin-Rubakov-Shaposhnikov electroweak baryogenesis.

\emph{Acknowledgements:} This work has been supported by the European Research Council (ERC) under the European Union's Horizon 2020 research and innovation programme (Grant Agreement No. 694248).

\end{document}